\documentclass{raa}            

\usepackage{mathrsfs}
\usepackage{graphicx,epstopdf,times}
\usepackage{natbib}
\usepackage{amssymb,amsmath}
\bibpunct{(}{)}{;}{a}{}{,}

\usepackage[pagebackref=true]{hyperref}
\newcommand{\subfigimg}[3][,]{%
  \setbox1=\hbox{\includegraphics[#1]{#3}}
  \leavevmode\rlap{\usebox1}
  \rlap{\hspace*{25pt}\raisebox{\dimexpr\ht1-2\baselineskip}{#2}}
  \phantom{\usebox1}
}

\graphicspath{{./}{figs/}}

\begin{document}

\title{X-ray fine structure of a limb solar flare revealed by Insight-HXMT, RHESSI and Fermi}


 \volnopage{ {\bf 20XX} Vol.\ {\bf X} No. {\bf XX}, 000--000}
   \setcounter{page}{1}

   \author{Ping Zhang
   \inst{1,2}, Wei Wang\inst{1,2},Yang Su\inst{4}, Shuangnan Zhang \inst{3}, Liming Song\inst{3}, Fangjun Lu\inst{3}, Shu Zhang \inst{3}
   }

   \institute{ Department of Physics and Technology, Wuhan University, Wuhan 430072, China; {\it wangwei2017@whu.edu.cn}\\
        \and
             WHU-NAOC Joint Center for Astronomy, Wuhan University, Wuhan 430072, China\\
        \and
            Institute of High Energy Physics, Chinese Academy of Sciences, Beijing 100049, China\\
        \and
            Purple Mountain Observatory of the Chinese Academy of Sciences, Nanjing, China \\
\vs \no
   {\small Received 20XX Month Day; accepted 20XX Month Day}
}






\abstract{
We conduct a detailed analysis of an M1.3 limb flare occurring on 2017 July 3, which have the X-ray observations recorded by multiple hard X-ray telescopes, including Hard X-ray Modulation Telescope (Insight-HXMT), Ramaty High Energy Solar Spectroscopic Imager (RHESSI), and The Fermi Gamma-ray Space Telescope (FERMI). Joint analysis has also used the EUV imaging data from the Atmospheric Imaging Assembly (AIA) aboard the Solar Dynamic
Observatory. The hard X-ray spectral and imaging evolution suggest a lower corona source, and the non-thermal broken power law distribution has a rather low break energy $\sim$ 15 keV. The EUV imaging shows a rather stable plasma configuration before the hard X-ray peak phase, and accompanied by a filament eruption during the hard X-ray flare peak phase. Hard X-ray image reconstruction from RHESSI data only shows one foot point source. We also determined the DEM for the peak phase by SDO/AIA data. The integrated EM beyond 10 MK at foot point onset after the peak phase, while the $>$ 10 MK source around reconnection site began to fade. The evolution of EM and hard X-ray source supports lower corona plasma heating after non-thermal energy dissipation. The combination of hard X-ray spectra and images during the limb flare provides the understanding on the interchange of non-thermal and thermal energies, and relation between lower corona heating and the upper corona instability.  
}


\authorrunning{P. Zhang, W. Wang et al. }            
\titlerunning{X-ray fine structure of limb flare }  

\maketitle

\section{Introduction} \label{sec:intro}
Solar corona is formed by hot plasma with complex magnetic filed. The super hot plasma formation is still a mystery in solar physics, and various processes are proposed based on alternating current (AC; \citealt{1947MNRAS.107..211A}) and direct current (DC;\citealt{1986GApFD..35..277P}) heating mechanisms. It is essentially that the heating originates from different types of energy conversion and accompanying response of the corona plasma, then studies address that the highly disparate spatial scales, physical connections between the corona and lower atmosphere, complex microphysics, variability and  dynamics could be the possible approaches for solving corona plasma heating, especially the physical connections between the corona and lower atmosphere (\citealt{2004psci.book.....A,2006SoPh..234...41K}). In fact, all those approaches are deeply rooted in the evolution of corona dynamics. And then the flare is one of the most energetic events occurred in the solar corona, which induced plasma heating, particle acceleration and multi-wavelength radiation (\citealt{2017LRSP...14....2B}), might provide a clear figure of energy conversion and plasma response. 

A general concept of the flare standard model believes that most of the eruptive energy comes from  magnetic reconnection in the solar corona. As the standard model (CSHKP; \citealt{ 1964NASSP..50..451C,1966Natur.211..695S,1974SoPh...34..323H,1976SoPh...50...85K}) described, a cusp-shaped structure with hot loop top and foot-points releases enormous energy through magnetic reconnection, with plenty of energy dissipation by various processes, such as plasma thermal conduction, which could been seen in soft X-rays and EUV; or accelerating non-thermal particles, then hard X-ray footpoint sources present, higher energy electrons also could be trapped in magnetism plasma loops and produce impressive radio emissions. In standard scenario high energy electrons propagates from higher corona into the dense chromosphere along complex magnetic loop (\citealt{2017LRSP...14....2B}).
It's natural that if we could obtain the precise 3D morphology structure of the flare and its evolution, which could be very helpful for better modeling of the physic process, determinate the energy budget and its conversion efficiency (\citealt{2021ApJ...913...97F}). But almost all ground and space observations only provide line of sight in field of view, projection effect becomes very important, which causes discrepancy between the observations and theory  (\citealt{1996ApJ...459..330F}), especially for detailed energy release and dissipation processes (\citealt{2016A&A...588A.116W}). 

Even though it's very difficult to avoid the projection effect from in situ and remote sensing observations, limb flare has the relatively distinct advantage. Limb flare usually provides a better side view of the flare morphology, especially the cusp-shaped structure evolution. Some researchers  \citep{2019NatCo..10.2276C,2020Sci...367..278F,2021ApJ...908L..55C} provided clear evidence revealing the location of the flare magnetic reconnection site, its current sheet formation and corona magnetic field evolution by utilizing the excellent limb super flare based on jointly hard X-ray, radio and ultraviolet observations. Especially the hard X-ray observations could reveal non-thermal electrons acceleration site, distribution and evolution in the hot plasma (\citealt{2019ApJ...877..145L,2021ApJ...918...42Z}), nevertheless super flares emitting bulk of photons usually result in instrument saturation, especially in X-ray and EUV bands, however, which are often used to determine the hot plasma characteristics.

For the past decade, the Atmospheric Imaging Assembly (AIA; \citealt{2012SoPh..275...17L}) provided an advance of high spatial resolution UV-visible/extreme ultraviolet (EUV) image data. Numbers of plasma differential emission measures (DEMs) method developed (\citealt{2004IAUS..223..321W,2012A&A...539A.146H,2015ApJ...807..143C,2018ApJ...856L..17S}), excavated the thermal plasma dynamics from few MK to tens of MK, which also very depended on the data quality. However during the flare period, thermal energy could not exist without non-thermal energy, and they convert to one another during different flare phases, so we also need hard X-ray imaging spectroscopy for better dynamic range of plasma characteristic determination, especially for gamma-ray flares in which non-thermal energy dissipation dominates (\citealt{1976SoPh...50..153L,2011SSRv..159..421L}), or super-hot flares in which thermal energy dissipation dominates (\citealt{2010ApJ...725L.161C,2014ApJ...788L..31C}).

In this study, we present the hard X-ray timing properties of an M1.3 limb flare by using observations from RHESSI, FERMI and Insight-HXMT, and investigate detailed hard X-ray spectral evolution with both RHESSI and FERMI data, meanwhile study the plasma loop evolution by SDO/AIA data. We shows the hard X-ray light-curves observed from RHESSI, FERMI and HXMT/CsI detectors, and the limb flare did not have too much high energy photons above 100 keV. And the hard X-ray spectral analysis shows a typical soft–hard–soft broken power law distribution, with a power law index around 3 at peak time. However, the hard X-ray image only shows foot-point structure. In the next section, the hard X-ray observations of the limb flare by different missions are introduced. In section 3, the detailed analysis of the spectral evolution of the flare in hard X-rays is described. The conclusion is presented in the last section.

\section{Hard X-ray observations} \label{sec:obs}

The Reuven Ramaty High Energy Solar Spectroscopic Imager (RHESSI) is a NASA Small Explorer Mission (\citealt{2002SoPh..210....3L}). It could resolve the hard X-ray source on the solar disk with 4 s high cadence, could make imaging and spectroscopy measurements from soft X-rays to gamma-rays (i.e., 3 keV to 17 MeV). The highest angular resolution is $\sim $3 $\arcsec$, and the highest energy resolution is down to 1 keV. Gamma-ray Burst Monitor (GBM) on board the Fermi Gamma-ray Space Telescope consists of an array of 12 thallium-doped sodium-iodide detectors (NaI(Tl)) (\citealt{2009ApJ...702..791M}), covering an energy range of 8 keV to 1 MeV, and the detectors also have the capacity to detect incident solar hard X-ray and gamma-ray photons.
The Insight-HXMT satellite consists of three main telescopes (\citealt{2020SCPMA..63x9502Z}): High energy telescope (HE), Medium Energy telescope (ME) and Low Energy telescope (LE) . The HE telescope has 18 cylindrical NaI/CsI detectors, the NaI detector could resolve 20-250 keV sources in the field of view, besides the CsI detector could detect gamma rays from 80-800 keV (Normal-Gain mode) and 200-3000 keV (Low-Gain mode) respectively (\citealt{2020SCPMA..6349503L,2020JHEAp..27....1L}). The CsI detectors also could detect incident solar hard X-ray and gamma ray photons (\citealt{2021ApJ...918...42Z}). The Solar Dynamic Observatory Atmospheric Imaging Assembly(SDO/AIA; \citealt{2012SoPh..275....3P}) could resolve the flare loops and plasma from 0.02 MK to more than 20 MK, with quite high spatial resolution $\sim$ 0.6 $\arcsec$, and the time resolution $\sim$ 12 s. In this study we will use SDO/AIA data for DEM inversion and only use 131 $\AA$ for flare region imaging, which is sensitive to hot flare plasma in the corona. 

\begin{figure}[hb!]
\centering
\includegraphics[width=12.0cm, angle=0]{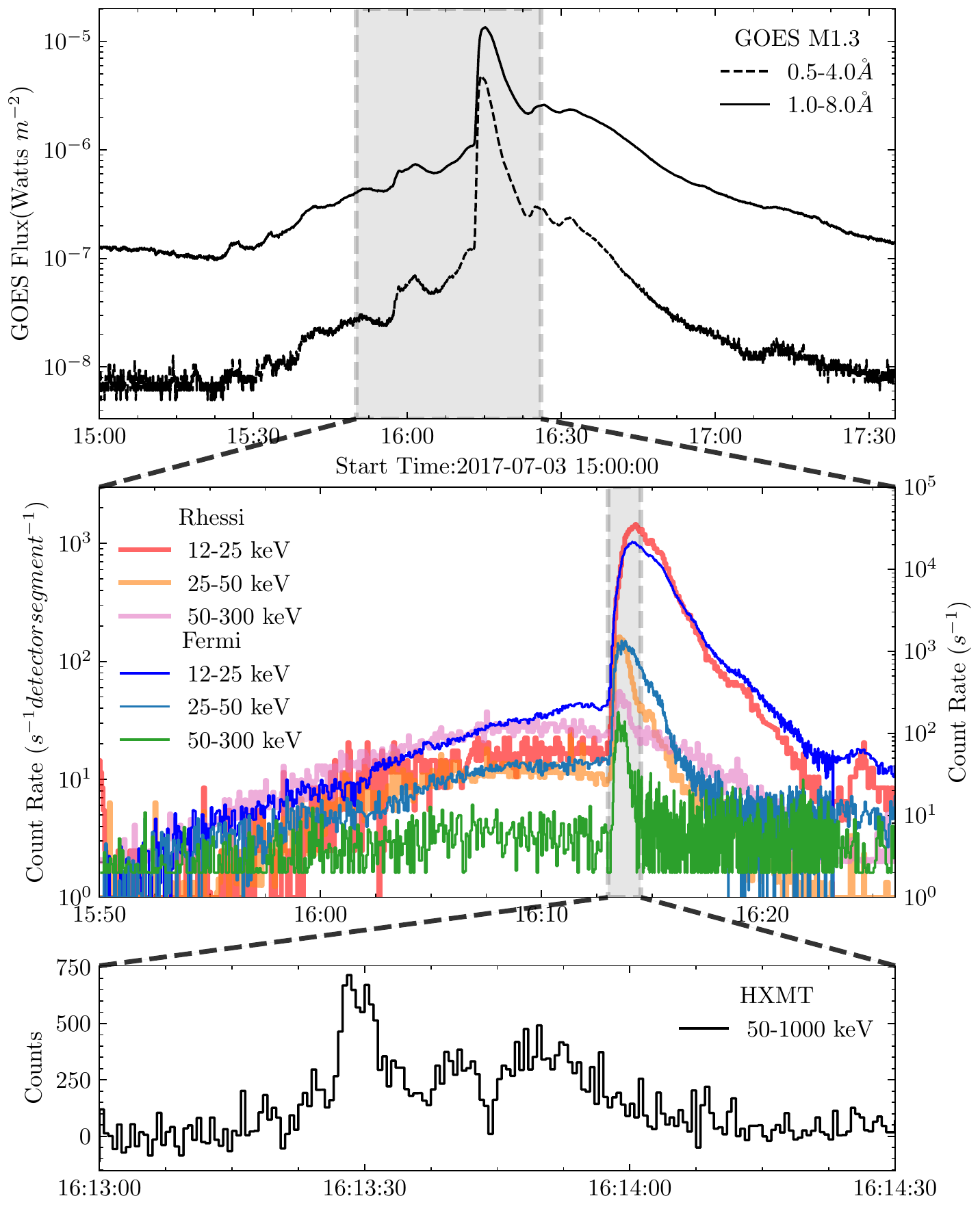}
\caption{The multi-wavelength light curves of the M1.3 flare: (upper panel) the GOES soft X-ray light curves, the dashed line is $0.5-4.0 \AA$, solid line is $1.0-8.0 \AA$; (middle panel)  the hard X-ray light curve, red orange and pink solid lines represent RHESSI observations and blue, light blue and green solid lines stand for FERMI observations;( bottom panel) light curve observed by Insight-HXMT CsI detectors at 50-1000 keV range. \label{fig:lc}}
\end{figure}

The limb M1.3 class flare occurred at 2017 July 3 15:37 UT, with the X-ray light curves as shown in Figure \ref{fig:lc}. We could see the flare lasted for almost 1.5 hours in the soft X-ray band from GOES observations. However the hard X-ray photons were detected only near the flare peak phase, and the onset of hard X-rays was from 15:50 UT according to RHESSI and FERMI observations. In Figure \ref{fig:lc}, we also present the Insight-HXMT CsI light curve in the peak phase. In the middle panel we could see that there were very limited photons higher than 50 keV, FERMI and RHESSI light curves well matched during the whole flare phase. However, since Insight-HXMT CsI detector have higher threshold for gamma-ray photons \citep{2020JHEAp..27....1L}, we only see the excess of the count rates at peak phase. 
   
\section{Hard X-ray evolution characteristics} \label{sec:results}

\begin{figure}[ht!]
  \begin{minipage}[t]{0.32\linewidth}
  \centering
   \subfigimg[width=50mm]{\bf a}{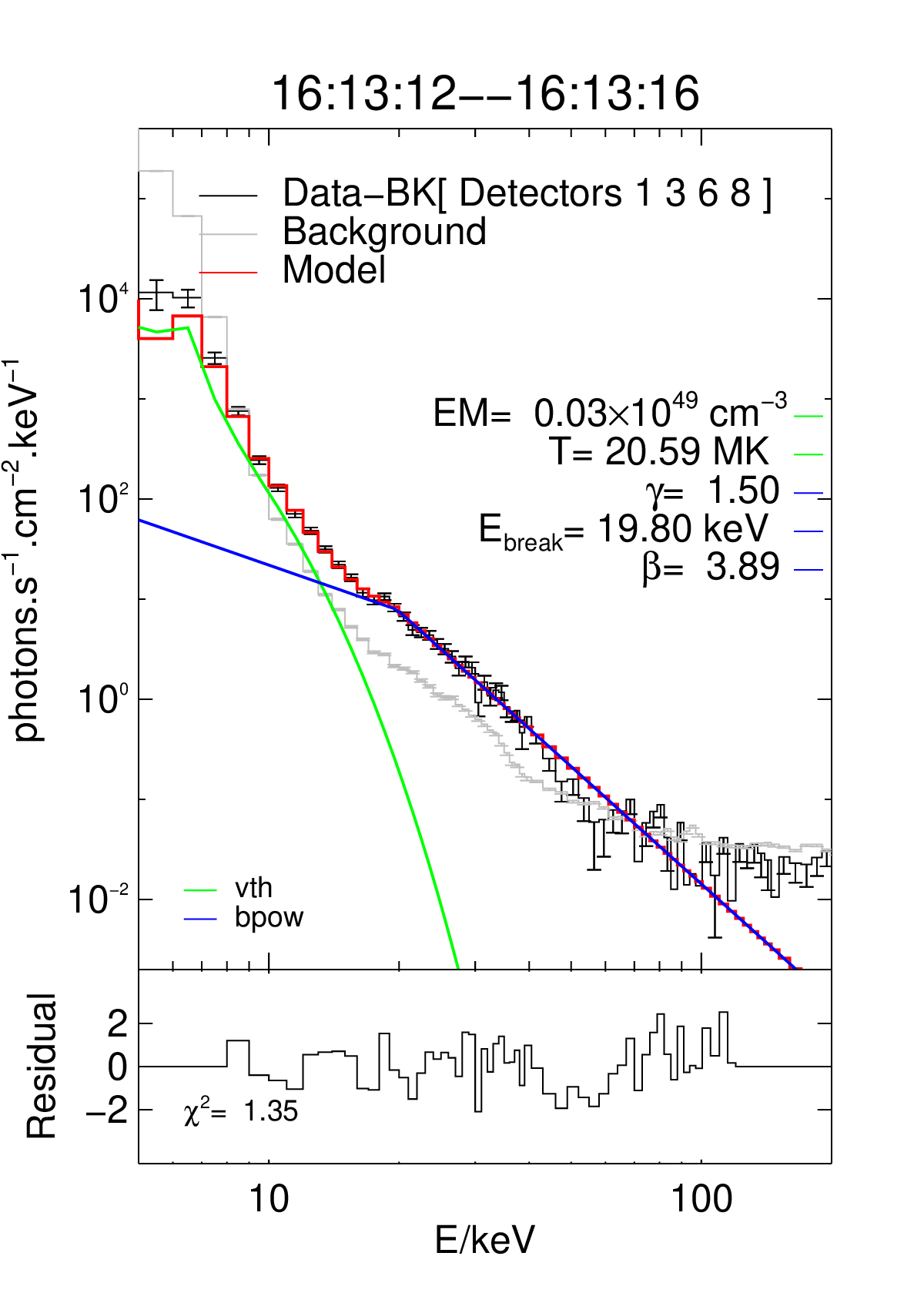}
  \end{minipage}%
  \begin{minipage}[t]{0.32\textwidth}
  \centering
   \subfigimg[width=50mm]{\bf b}{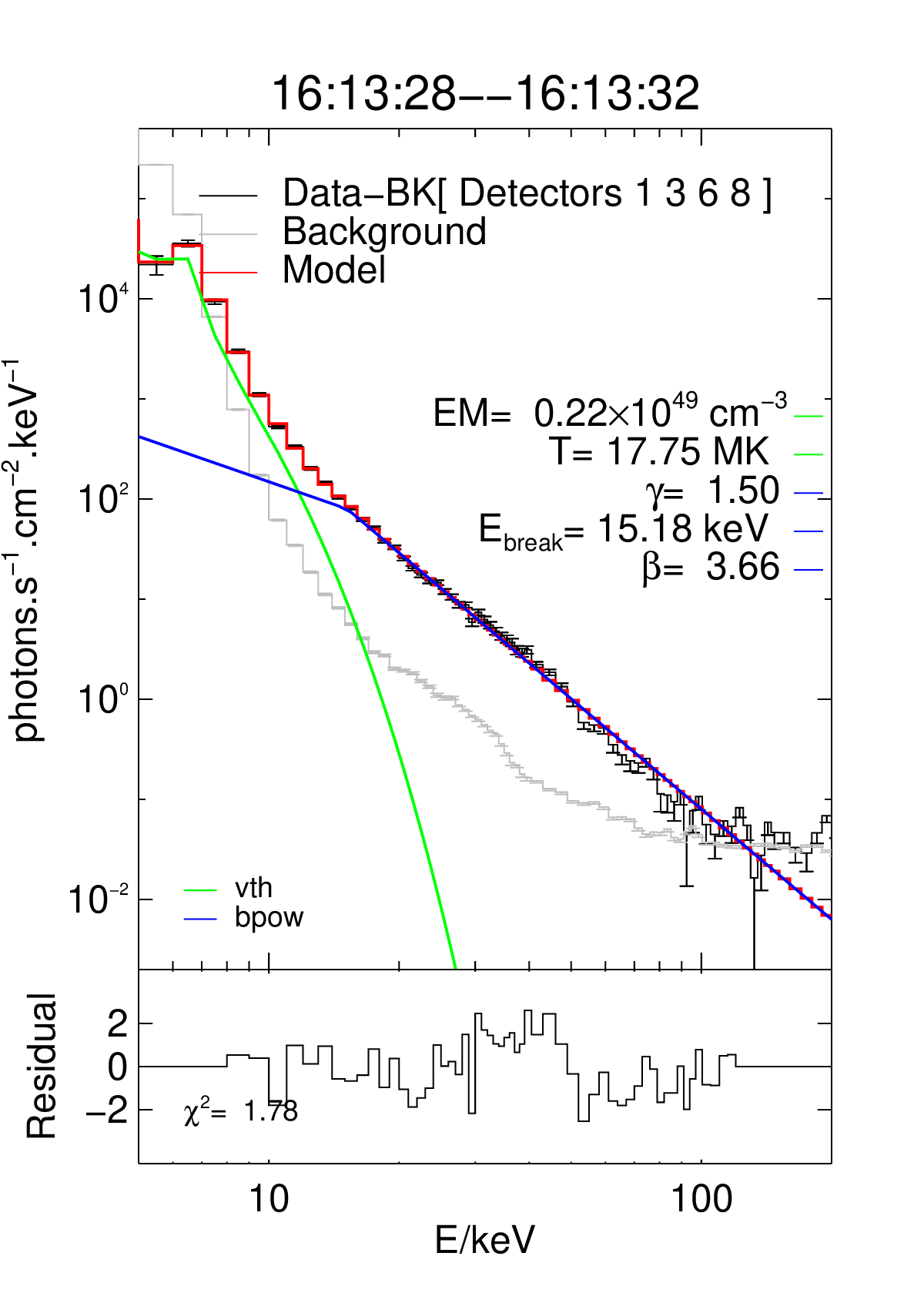}
  \end{minipage}%
    \begin{minipage}[t]{0.32\textwidth}
  \centering
   \subfigimg[width=50mm]{\bf c}{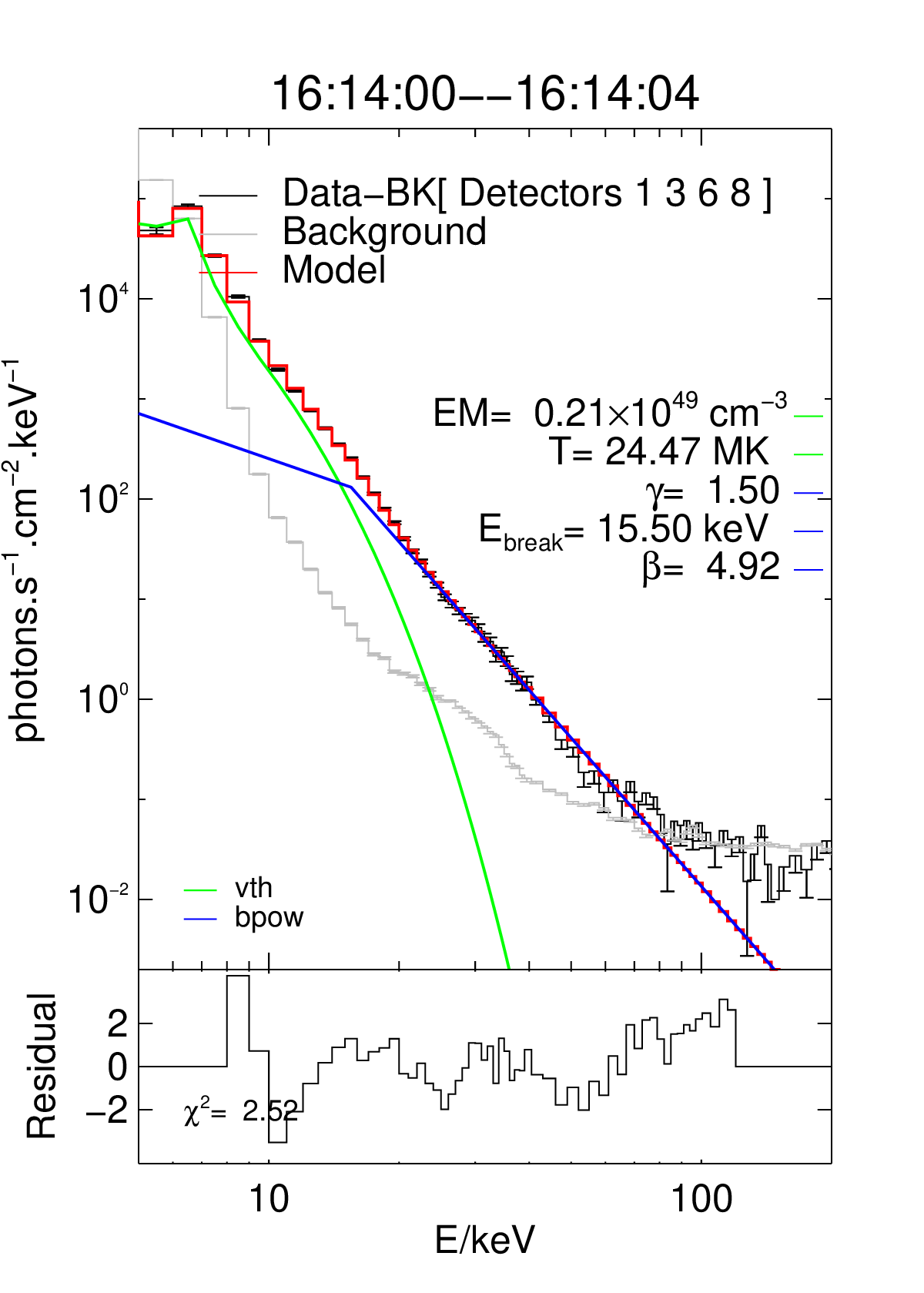}
  \end{minipage}%
    \caption{The figures show the hard X-ray spectra and their spectral fitting results based on RHESSI observations before the hard X-ray peak phase ({\bf a}, 16:13:12-16:13:16 UT), at peak time ({\bf b},16:13:28-16:13:32 UT) and after the hard X-ray peak phase ({\bf c},16:14:00-16:14:04 UT). The green solid lines represent the thermal component, while the blue solid lines are the non-thermal component with the broken power-law model.}
    \label{fig:spec}
\end{figure}

Both RHESSI and FERMI/GBM have full observations of the limb flare, and their spectral features were consistent with each other in the range of $\sim 10 -100$ keV. In this study , we only analyzed the RHESSI spectral properties and the variations around the peak phase time in hard X-ray bands as shown in Figure \ref{fig:lc}. The hard X-ray spectra could be well fitted with thermal plus broken power law model. We have shown the spectra examples for three typical time intervals in Figure \ref{fig:spec}: before the peak, at the peak and after the peak. The narrow peak time interval was defined around 16:13:00 - 16:14:30 UT based on the hard X-ray light curve obtained by Insight-HXMT (see Figure \ref{fig:lc}). Figure \ref{fig:spec}(a) shows the thermal plasma tended to be very hot of the temperature $\sim$ 20 MK, and photons above 50 keV almost at the background level before the hard X-ray peak. However bulk of energy would start to accelerate electrons in the corona, and plenty of thermal electrons gaining sufficient energy turn into non-thermal electrons, which are the origin emitter of hard X-ray photons and centimeter-wavelength radio waves. At the hard X-ray curve peak, we could see that the non-thermal power law index became harder while the hot plasma became cooler with $T\sim$ 17 MK as shown in Figure \ref{fig:spec}(b). Then after the peak phase as shown in Figure \ref{fig:spec} (c), the energetic electrons lose all their energy in the dense corona, which in addition continue to heat the corona source, so that we got the hot plasma of $T\sim $ 24 MK but softer power law distribution at non-thermal region.

According to the light curves shown in Figure \ref{fig:lc}, we divided the whole peak phase covering the time interval from 16:13:12 UT to 16:16:30 UT into several small time intervals, and obtained the hard X-ray spectra for each time interval. Then we continue the spectra fittings with RHESSI data for each time interval through the flare peak phase from 16:13:12 UT to 16:16:30 UT with the same model described above.

\begin{figure}[ht!]
\centering
\includegraphics[width=12.0cm, angle=0]{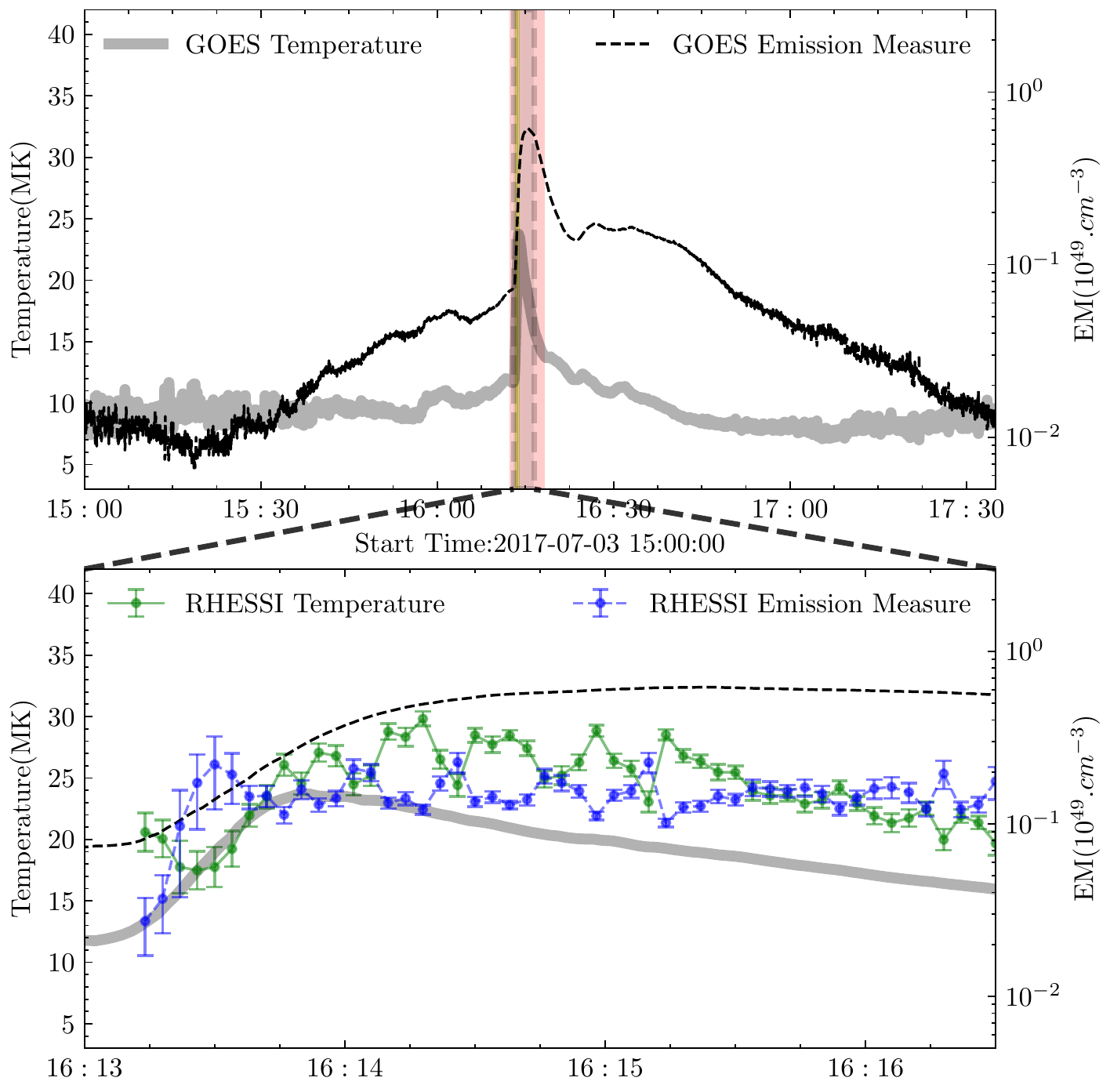}
\caption{The EM and temperature evolution derived from GOES and RHESSI. Top panel: plasma temperature and EM derived from GOES data, and the RHESSI spectral fitting range marked with pink; Bottom panel: the green and blue curves with error bars stand for plasma temperature and EM evolution derived from RHESSI spectral fittings.  } \label{fig:emcom}
\end{figure} 

The thermal plasma configuration contributes most soft X-ray emission, according to \citet{1994SoPh..154..275G} and \citet{2005SoPh..227..231W}, under the isothermal assumption one could derive the plasma emission measure and temperature via GOES fluxes in the two channels based on CHIANTI database with both coronal and photospheric abundance models. In Figure \ref{fig:emcom}, the plasma characteristics derived from GOES observations were demonstrated in the top panel, and in the bottom panel we compared the plasma emission measure and temperature evolution obtained from the RHESSI spectral fitting during the whole peak phase. During the peak phase, the temperature of the hot plasma source emitting hard X-rays tended to decrease before 16:13:32 UT and started to increase after the peak, then slowly decreased after 16:14:30 UT On the contrary, RHESSI emission measure started to increase before 16:13:32 UT and slowly decreased (nearly in the stable value level) later. It is consistent with the corona plasma configuration that higher temperature source tends to have a smaller emission volume and lower temperature source tends to have a bigger emission volume. It should be noted that the plasma configuration difference between GOES and RHESSI was attributed to the different energy ranges (\citealt{2014SoPh..289.2547R}).

\begin{figure}[ht!]
\centering
\includegraphics[width=12.0cm, angle=0]{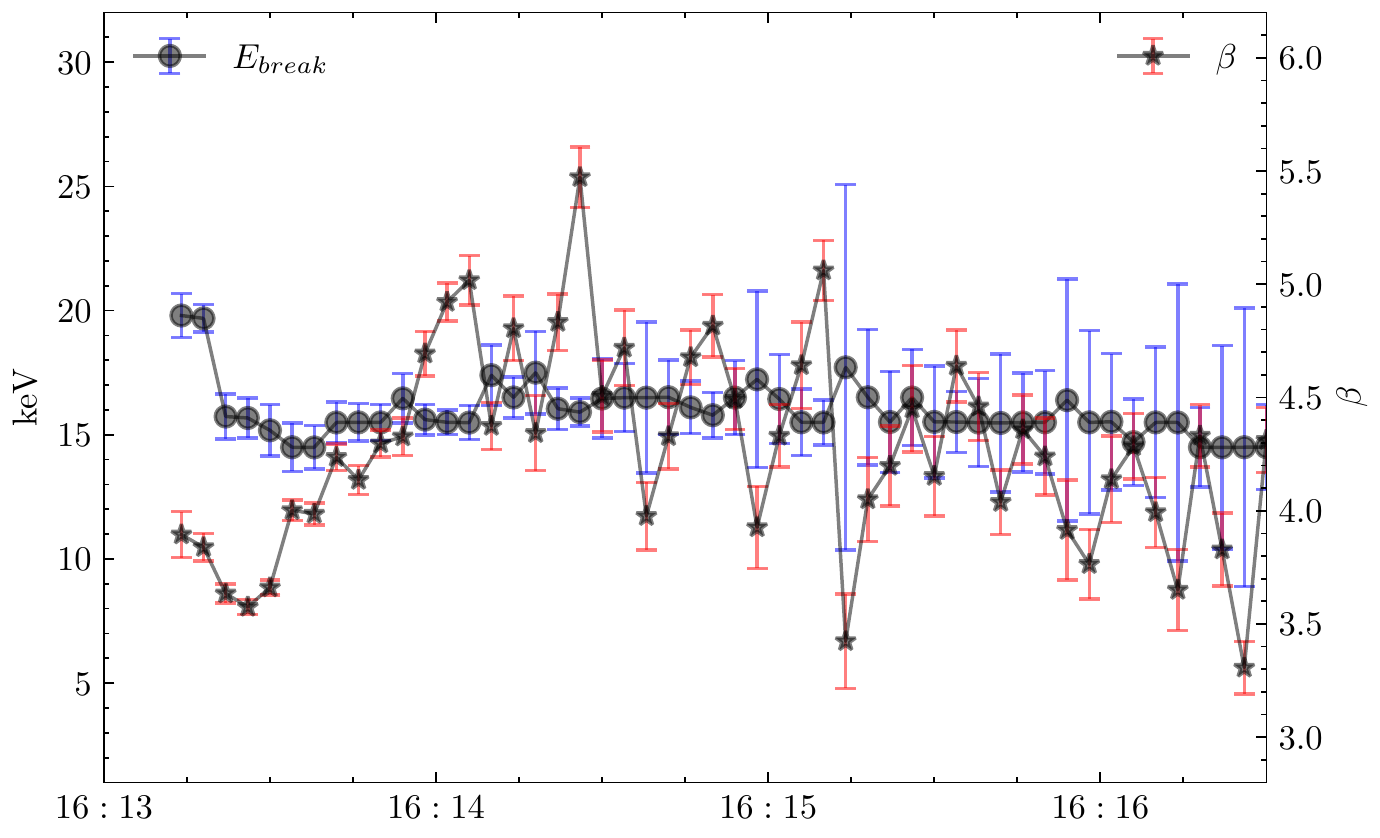}
\caption{The beta and break energy evolution during the peak phase derived from RHESSI. Blue circle represents the break energy in the broken power law fitting; red star stands for the power law index $\beta$ above the break energy. } \label{fig:nont}
\end{figure}  

The non-thermal component evolution during the peak phase were also shown in Figure \ref{fig:nont}. The break energy during the peak phase was around 20 keV before the peak (two data points), and decreased to be $\sim$ 15 keV with very minor variation after then. On the other way, power law index $\beta$ has shown a soft-hard-soft variation pattern from 3.5 to 5.5, which is also consistent with the flare model in previous studies. Both error bars tend to be bigger after 16:14:30 UT might due to the non-thermal energy dissipation which leads to smaller non-thermal photons comparable with background.     

\begin{figure}[htb!]
\centering
\includegraphics[width=15.0cm, angle=0]{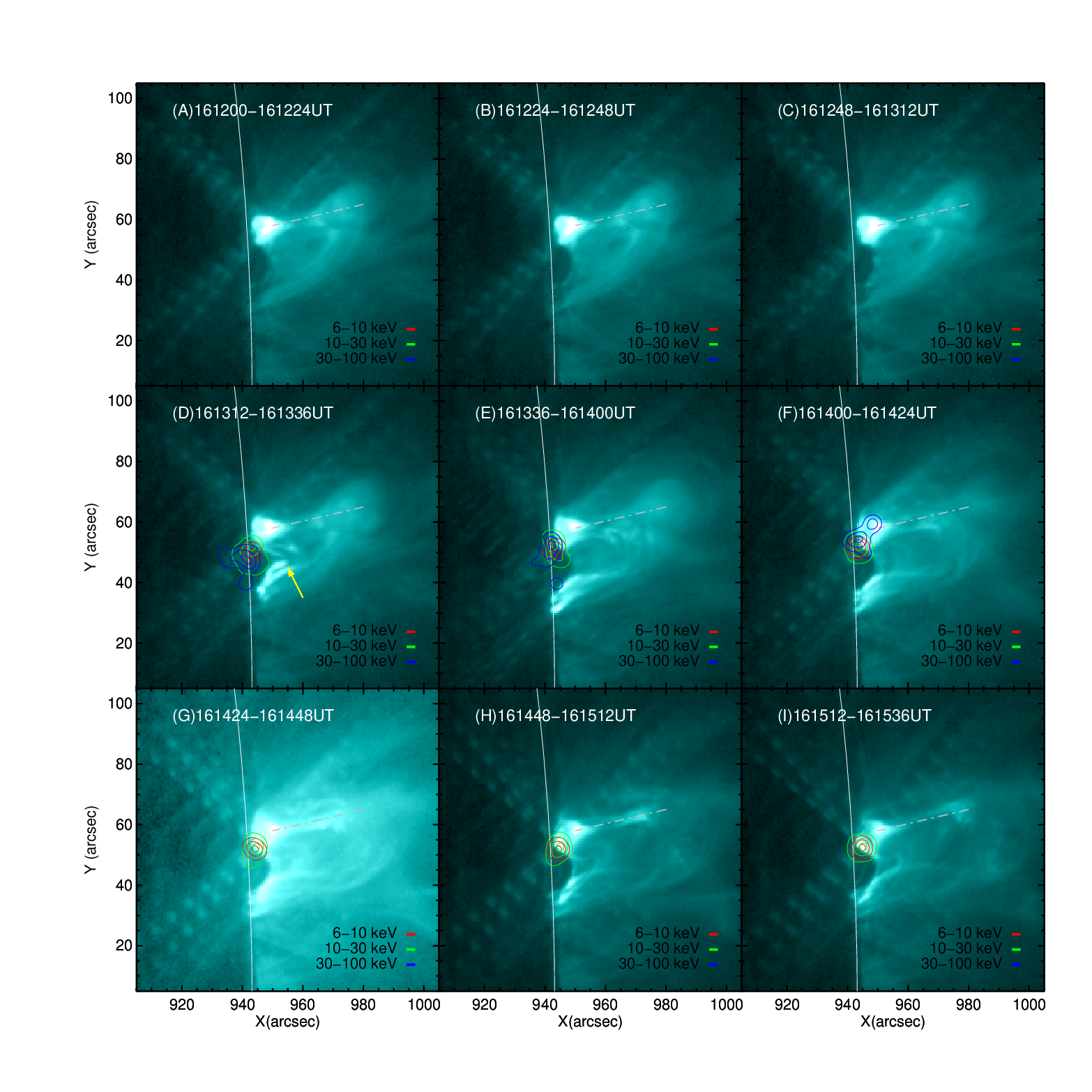}
    \caption{The figure shows the composite images using data from SDO/AIA 131$\AA$ observations in 9 integration time ranges from 16:12:00 to 16:15:36 UT, solar limb marked with gray line, the integration time is 24 seconds. We also overlay the contours of hard X-ray images based on RHESSI data in three energy bands: 6-10 keV, 10-30 keV and 30-100 keV marked with red, green and blue solid lines respectively. The contour levels for 6-10 keV, 10-30 keV and 30-100 keV are 30$\%$, 60$\%$ and 90$\%$. The gray dash-dotted lines mark the reconnection current sheet. The yellow arrow in panel (D) marks the first appearance of the erupting filament.}
    \label{fig:euv}
\end{figure}

To locate the hard X-ray emission site, we have used RHESSI data to reconstruct the hard X-ray image at the flare region. Besides the hard X-ray imaging data, we also used the UV observations from Atmosphere Imaging Assembly (AIA). In Figure \ref{fig:euv}, we shows the EUV color images with AIA wavebands 131 $\AA$ (which is very sensitive for million Kelvin plasma at flare region) in 9 time intervals around the flare peak phase from 16:12:00 UT to 16:15:36 UT, with overlaying the hard X-ray contours based on RHESSI observations on the SDO/AIA data. The red, green and blue contours represent three hard X-ray bands: 6-10 keV, 10-30 keV and 30-100 keV respectively, and we used different contour levels for higher energy ranges because at higher energies the imaging quality is not as good as lower bands. However the hard X-ray images did not show complex morphology structure compared with EUV data, we only found a very simple condensed hard X-ray foot-point at the bottom of hot plasma loops. It should be clarified that in order to increase the signal to noise ratio, we reconstructed RHESSI data with 24 second integration at 6 time intervals from 16:13:12 UT to 16:15:36 UT using CLEAN algorithm\citep{2002SoPh..210...61H}. Given that hard X-ray image reconstruction should have sufficient photons in specific energy band, we have put $10^4$ total counts after reconstruction as the threshold. 

We could see a fine stable plasma filament before 16:13:12 UT in Figure \ref{fig:euv}(A/B/C), and a mini-cusp structure arose at the peak of hard X-ray phase as yellow arrow pointed in Figure \ref{fig:euv} D, which was later accompanied by a filament eruption. We also mark the current sheet with the gray dot line, then one could see the current sheet survival through the whole hard X-ray flare phase. The most brightening area at the bottom of the current sheet should be the flare reconnection site. Figures \ref{fig:euv}(D to I) show that the hard X-ray source is mainly located at the northern foot-point of EUV flare loops. The filament eruptions started and were accompanied by the hard X-ray flare peak phase after 16:13:12 UT. The filament eruption might mainly be caused by the upward motion of the overlying loops. We did not see loop top hard X-ray source, but the brightening of 131 $\AA$ source and the dimming of high energy hard X-ray source at the foot point area, possibly because most non-thermal electrons lost their energy in the bottom corona via radiation process, and in the meantime particle collisions indirectly heat the bottom source.

\begin{figure}[htb!]
\centering
\includegraphics[width=15.0cm, angle=0]{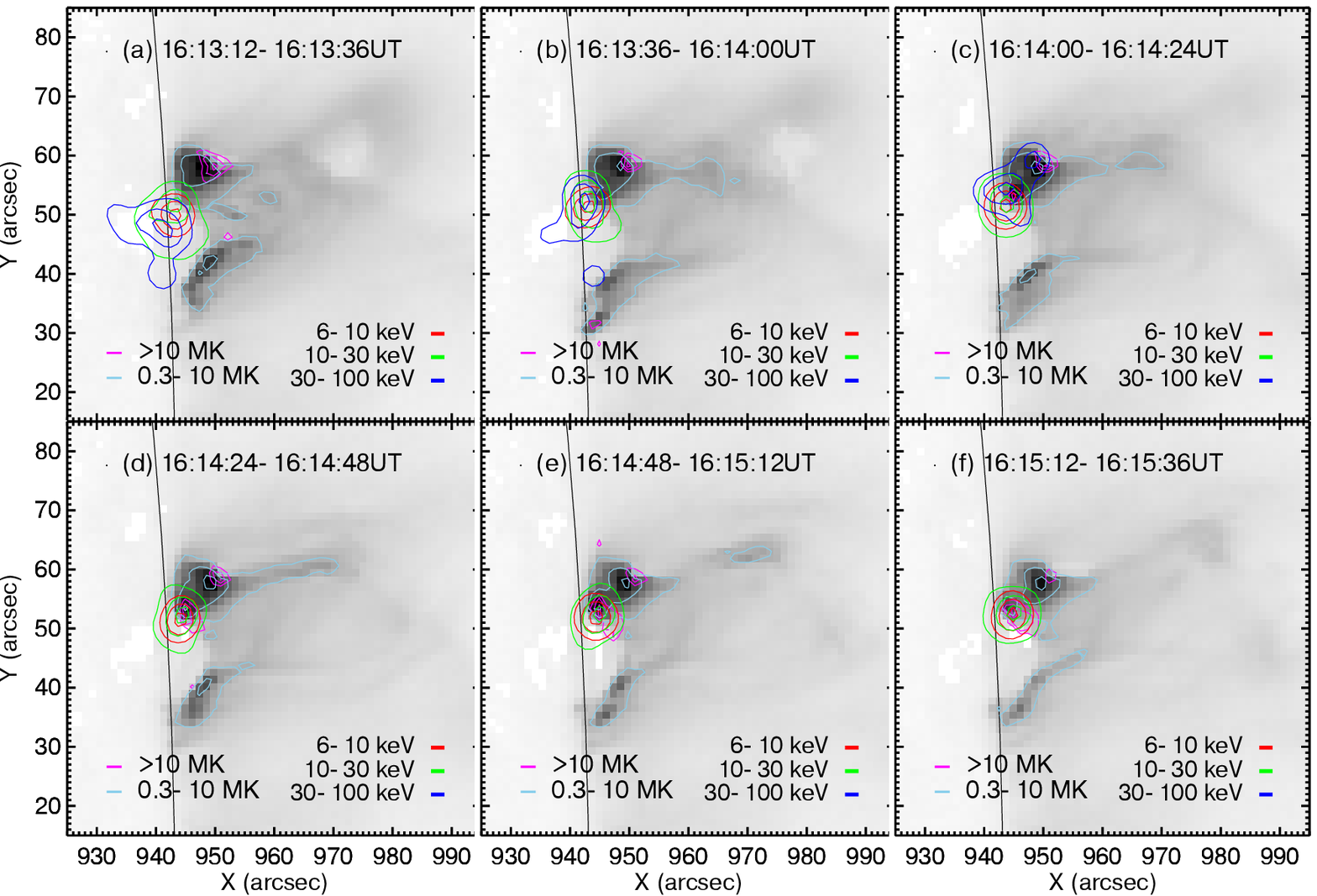}
    \caption{The figure shows the EM maps derived from SDO/AIA  observations in six integration time ranges from 16:13:12 to 16:15:36 UT. Background is 0.3-10 MK integrated EM map, We overlay the contours of both 0.3-10 MK and  $> $ 10 MK integrated EM map with contour levels : 30$\%$, 60$\%$, 90$\%$, marked with sky blue and magenta respectively. The hard X-ray images contours based on RHESSI data in three energy bands: 6-10 keV, 10-30 keV and 30-100 keV marked with red, green and blue solid lines respectively. The contour levels for 6-10 keV, 10-30 keV and 30-100 keV are 30$\%$, 60$\%$ and 90$\%$.}
    \label{fig:euvem}
\end{figure}

In general standard flare models suggest that hard X-ray emissions have a looptop source and two footpoint sources (\citealt{1995ApJ...451L..83S}). During the limb flare, the X-ray light-curves (see Figure \ref{fig:lc}) showed a quite gentle and gradual flare event, which implied that the magnetic energy release and dissipation also turned out to be relatively slower than impulsive events. Then in such short period, non-thermal electrons could not be accelerated to higher energies and form hard X-ray loop-top sources, and possibly most energetic non-thermal electrons accelerated streaming downward to foot-point then lost all energy in the dense hot foot point plasma. After the flare peak phase, i.e., from 16:14:48 UT, the hard X-rays above 30 keV cannot be resolved by the RHESSI detector, which also suggested the high energy electrons lost the energies very fast after the peak. Such scenario also supported by emission measure maps derived from SDO/AIA data using method from \citealt{2018ApJ...856L..17S} as shown in figure \ref{fig:euvem}, the $>$ 10 MK plasma sources appears in the upper corona, and decreased along with the high energy hard x-ray source, later the arise in the hard X-ray flare footpoint. Single foot-point hard X-ray source shown in Figure \ref{fig:euv} might be due to projection effect or solar disk occultation because we only have line sight of view observations for the limb flare.

In addition, we can find the structure of 0.3-10 MK source was rather stable during the whole hard X-ray flare phase in Figure \ref{fig:euvem}. The hard X-ray flare magnetic reconnection site confirmed by AIA 131 $\AA$ in Figure \ref{fig:euv} also arose the $>$ 10MK plasma source but disappeared after peak phase. Later the upper corona $>$ 10 MK source started dimming and the foot point site $>$ 10 MK source occurred.  Hence we believe that the lower corona reconnection released vast of energy heating the ambient cooler plasma and accelerated electrons simultaneously, while the acceleration site was very close to the bottom of the dense corona where the hard X-ray source only lasted for a few minutes. But the dissipation energy fully injected the hot plasma loops, the unevenly heating at different corona heights resulted in the instability of the whole flare region loops after the flare peak.

\section{Conclusion and summary}\label{sec:model}
In this paper, we presented joint observations of a limb flare on 2017 July 3 by RHESSI, FERMI and Insight-HXMT and SDO/AIA to study the hard X-ray evolution during the flare peak phase. The hard X-ray light curves around the peak phase showed a very gradual variation pattern, indicating that the energy release from magnetic reconnection processes tended to be gentle compared with impulsive flares \citep{2018A&A...615A..48Z,2021ApJ...918...42Z}. The RHESSI hard X-ray imaging only showed a single foot-point, so we do not see the loop-top source and even another foot-point at all energy ranges which are predicted by a standard cups flare model. Missing of the other foot-point might be due to its occultation by the solar disk because the imaging data only provided us line-sight of view. The possible answer to the absence of loop-top source might be that most energetic electrons were accelerated \textbf{} stream downward but not upward, and consequently bremsstrahlung radiation could emit sufficient hard X-ray photons observed by the RHESSI detector.

The scenario is also supported by RHESSI spectral fitting results. The broken power law distribution gave the soft-hard-soft pattern in $\sim$ 30 s exactly around the hard X-ray peak time, but tended to have the gradual variation after peak phase. The same pattern appeared in the fluctuation of thermal plasma temperature, in addition we also see the inverted pattern of the emission measure evolution derived from RHESSI observation. The low break energy $\sim 15$ keV indicate that non-thermal electrons could not be accelerated to higher energies in this case, a considerable chunk of energy was still trapped in the corona loops, with rather a small population of electrons energization and stream downward to lower corona. But the dissipation of the non-thermal energy directly caused the fluctuation of thermal plasma configuration as shown in Figure \ref{fig:emcom} after the peak phase, and the fluctuation became weaker after 16:15:30 UT, when the loops at higher corona were broken. 

At present, different hard X-ray telescopes provide the insight to the particle accelerations and high energy emission properties in flares. \cite{2020ApJ...893L..40C} used the Nuclear Spectroscopic Telescope ARray (NuSTAR) to constrain the thermal plasma dynamics and the upper limits of non-thermal emission. In addition, the Insight-HXMT also presented the non-thermal X-ray/gamma-ray diagnose of a big flare, revealing the evolution of high energy electrons and corona magnetic field \citep{2021ApJ...918...42Z}. However, the absence of vivid hard X-ray source imaging poses challenges to current instruments on sensitivity and the angle of view. Recently the Spectrometer/Telescope for Imaging X-rays (STIX) on board Solar Orbiter present several cases of microflares observed during its instrument commissioning phase which provided better insights into the thermal and non-thermal energy dissipation during microflares (\citealt{2021arXiv210610058B}). Moreover, the Focusing Optics X-ray Solar Imager (FOXSI-2) sounding rocket experiment utilized a direct imaging technique  with impressive improvements in sensitivity and imaging dynamic range compared to RHESSI and demonstrated the presence of high temperature plasma $\sim$ 10 MK \citep{2016SPIE.9905E..0EG,2021ApJ...913...15V}.  Therefore, the joint X-ray observations of multiple instruments should be very helpful to understand the energy conversion and plasma response during the flares in wide energies and large spatial ranges. 
 

\begin{acknowledgements}
 We are grateful to the referee for the useful suggestions to improve the manuscript. This work is supported by the National Program on Key Research and Development Project (Grants No. 2021YFA0718500, 2021YFA0718503), the National Natural Science Foundation of China (Grants No. 12133007, U1838103, 11622326) and the Fundamental Research Funds for the Central Universities. This work made use of data from the \textit{Insight}-HXMT mission, a project funded by China National Space Administration (CNSA) and the Chinese Academy of Sciences (CAS).
\end{acknowledgements}

%

\vspace{5mm}







\bibliography{ms2021-0328}{}
\bibliographystyle{aasjournal}



\end{document}